\documentclass[referee]{mn2e}
\usepackage{graphicx}
\usepackage{subfigure}
\usepackage{longtable}
\usepackage{ltxtable}

\begin{document}

\newcommand{\etal}{et~al.}
\newcommand{\Msun}{\mbox{M}$_{\odot}$}
\newcommand{\WHz}{$\mbox{W\,Hz}^{-1}$}

\title[AGN and colours of local galaxies]
{AGN feedback drives the colour evolution of local galaxies}

\author[Shabala \etal\/]{Stanislav S. Shabala$^1$\footnote{Present address: School of Mathematics \& Physics, Private Bag 37, University of Tasmania, Hobart 7001, Australia}, Sugata Kaviraj$^{1,2}$ \& Joseph Silk$^1$ \\
$^1$ Oxford Astrophysics, Denys Wilkinson Building, Keble Road, Oxford OX1 3RH, United Kingdom\\
$^2$ Blackett Laboratory, Imperial College London, London SW7 2AZ, United Kingdom\\
Email: Stanislav.Shabala@utas.edu.au}

\maketitle

\begin{abstract}

We investigate the effects of AGN feedback on the colour evolution of galaxies found in local ($z<0.2$) groups and clusters. Galaxies located within the lobes of powerful Fanaroff-Riley type II (edge-brightened) sources show much redder colours than neighbouring galaxies that are not spatially coincident with the radio source. By contrast, no similar effect is seen near Fanaroff-Riley type I (core-dominated) radio sources. We show that these colours are consistent with FR-IIs truncating star formation as the expanding bow shock overruns a galaxy. We examine a sample of clusters with no detectable radio emission and show that galaxy colours in these clusters carry an imprint of past AGN feedback. AGN activity in the low-redshift Universe is predominantly driven by low-luminosity radio sources with short duty cycles. Our results show that, despite their rarity, feedback from powerful radio sources is an important driver of galaxy evolution even in the local volume.

\end{abstract}


\section{Introduction}
\label{sec:introduction}

There has been much recent interest in the properties of Active Galactic Nuclei (AGN) and the interaction between these objects and their host galaxies. Observationally, strong evidence supporting such interaction exists. Up to 70\% of cD galaxies contain radio sources at their centres \cite{Burns90}. These radio sources drive shocks into the intra-cluster medium (ICM), heating and transporting outward the cluster gas. The spatial coincidence of radio emission with cavities in the hot X-ray emitting gas \cite{FabianEA03,FormanEA05} agrees well with analytical and numerical models \cite{ChurazovEA01,BassonAlexander03}, and is required to explain the lack of catastrophic cooling expected in the absence of a feedback mechanism in dense cluster cores \cite{BinneyTabor95}.

Another piece of evidence comes from the field of galaxy formation. The growth of structure through the gravitational instability provides a simple and accurate description of the build-up of dark matter scaffolding through cosmic time, however AGN feedback is required to prevent runaway cooling from forming too many massive galaxies \cite{CrotonEA06,SA09b}. The tight correlation between galaxy and black hole masses \cite{MagorrianEA98,HaeringRix04} provides further evidence for the tight coupling between AGN and galaxy evolution.

Broadly speaking, two types of AGN are often discussed. Powerful outbursts observable at both optical and radio wavelengths are believed to be triggered by gas-rich mergers, facilitating a sudden influx of large quantities of fuel to the AGN. These are often observed in the local volume as Seyfert AGN, and as quasars at higher redshifts. They are almost invariably accompanied by strong radio emission. By contrast, the vast majority of radio-loud AGN at $z \sim 0$ do not have a corresponding Seyfert phase. The number counts of these objects are consistent with low-luminosity radio AGN activity being triggered by the cooling out of hot gas from the galactic halo \cite{BestEA05b,SAAR08}.

In a seminal work, Fanaroff \& Riley \shortcite{FR74} showed that the luminosity of a radio AGN is intrinsically tied to its morphology. The most powerful radio sources are invariably edge-brightened (Fanaroff \& Riley referred to these as type II objects, or FR-IIs), propagating bow shocks through the surrounding gas as they expand. The less powerful FR-I sources, on the other hand, have radio emission that is core-dominated. These objects dominate radio source counts in the local volume, and are typically interpreted as AGN that start out as FR-IIs but are not powerful enough to traverse the dense galaxy or cluster core without being disrupted by Rayleigh-Taylor and Kelvin-Helmholtz instabilities.

AGN feedback influences galaxy formation and evolution by affecting the gas available for star formation. Recently, Kaviraj \etal\/ \shortcite{KavirajEA11} showed that AGN feedback is required to explain the rapid transition of early-type galaxies from the blue cloud to the red sequence. There are a number ways in which AGN feedback can affect the gas. Kinetic feedback, typically associated with powerful radio sources, can shock heat and uplift the gas to large radii. The AGN radiation field can also heat the gas and drive outflows. Finally, turbulence induced by disruption of the jet-inflated radio cocoons, a process mostly associated with FR-I sources, can either trigger star formation (by aiding gravitational collapse; e.g. Silk \& Nusser 2010) or suppress it (by heating the gas further). The relative importance of these processes is unclear. In particular, it is not clear how important feedback from the powerful but rare FR-IIs is compared with the less powerful but omnipresent FR-Is. We aim to address this issue in the present paper.

The paper is structured as follows. In Section~\ref{sec:sample} we describe our sample. Results are presented in Section~\ref{sec:results}, and discussed in Section~\ref{sec:discussion}. We conclude in Section~\ref{sec:summary}.

\section{Sample selection}
\label{sec:sample}

In this section we describe the construction of a sample of galaxies around radio sources.

\subsection{AGN selection}
\label{sec:agnCatalogues}

The radio source list was compiled from a combination of catalogues. The largest radio AGN in terms of angular size were selected from the 3CRR sample of Laing \etal\/ \shortcite{LRL83}\footnote{Leahy, Bridle \& Strom provide a nice compilation of these objects at www.jb.man.ac.uk/atlas}, and complemented by additional data from Machalski \etal\/ \shortcite{MachalskiEA04}. Smaller radio sources were identified from two large-area 1.4~GHz surveys, Faint Images of the Radio Sky at Twenty Centimetres (FIRST; Becker \etal\/ 1995) and the NRAO VLA Sky Survey (NVSS; Condon \etal\/ 1998). These are complementary in the sense that FIRST has higher sensitivity, but NVSS is better at picking up extended objects. Shabala \etal\/ \shortcite{SAAR08} identified optical Sloan Digital Sky Survey (SDSS; Strauss \etal\/ 2002) counterparts for FIRST/NVSS radio sources to $z=0.1$. We limit the present work to $z<0.2$, which provides a good balance between statistically meaningful number counts of galaxies and relatively low contamination in radio/optical positional matching due to line-of-sight effects (which increases with redshift). For $0.1 \leq z < 0.2$, we selected radio sources from the CoNFIG sample \cite{GW09}.

This approach allows construction of a sample containing both the more powerful, edge-brigtened FR-II sources, as well as the less powerful edge-darkened FR-Is. Details are given in Tables~\ref{tab:FR2s} and \ref{tab:FR1s}. We note that while the sample is not complete, we expect it to be representative as it spans a wide range of radio source sizes, luminosities, morphologies and ages.

\begin{longtable}{lllllll}
%\centering
%\begin{tabular}{lcccccc}
%\hline
%\multicolumn{1}{l}{IAU} & Other & redshift & LAS & Axial & $t_{\rm syn}$ \\ 
%\multicolumn{1}{l}{name} & name & & (arcsec) & ratio & (Myr) \\
%\hline
\hline
IAU & Other & redshift & frequency & LAS & Axial & $t_{\rm syn}$ \\ 
name & name & & (GHz) & (arcsec) & ratio & (Myr) \\
\hline
\endhead
\multicolumn{2}{l}{}\\
\hline
IAU & Other & redshift & frequency & LAS & Axial & $t_{\rm syn}$ \\ 
name & name & & (GHz) & (arcsec) & ratio & (Myr) \\
\hline
\endfirsthead
\multicolumn{2}{l}{{\it continued on next page}}
\endfoot
\multicolumn{2}{l}{}
\endlastfoot
0010-1108	& 	& 0.077	& 1.4 & 200	& $4^c$	& \\
0057-0052	& 	& 0.044	& 1.4 & 17	& $3.7^c$	& \\
0131+0033	& 	& 0.079	& 1.4 & 46	& $5.2^c$	& \\
0739+3947	& 	& 0.098	& 1.4 & 33	& $4^c$	& \\
0745+3357	& 	& 0.063	& 1.4 & 14	& $3.5^c$	& \\
0756+3703	& 	& 0.077	& 1.4 & 48	& $5^c$	& \\
0758+3747	& 	& 0.041	& 1.4 & 43	& $3.8^c$	& \\
0805+2409	& 3C192	& 0.060	& 1.41 & 201	& $4.33^a$      & \\
0819+5232	& 	& 0.189	& 1.4 & 49	& $1.6^d$	& \\
0821+4702	& 	& 0.130	& 1.4 & 38	& $1^d$	& \\
0902+5203	& 	& 0.099	& 1.4 & 28	& $2.2^c$	& \\
0911+3724	& 	& 0.105	& 1.4 & 60	& $2.2 \pm 0.2^b$	& $28 \pm 5^b$ \\
0921+4538	& 3C219	& 0.174	& 1.52 & 190	& $2.544^a$	& $60^f$	 \\
0930+0348	& 	& 0.089	& 1.4 & 33	& $5.4  ^c$	&  \\	
0939+3553	& 3C223	& 0.138	& 1.50 & 306	& $3.926^a$	& $72^g$	 \\
0941+3944	& 	& 0.107	& 1.4 & 190	& $1.8  ^d$	&  \\	
0947+0725	& 	& 0.087	& 1.4 & 310	& $5.5  ^d$	&  \\	
0949-0050	& 	& 0.081	& 1.4 & 25	& $2.4  ^c$	&  \\	
0955+0135	& 	& 0.099	& 1.4 & 44	& $4    ^c$	&  \\	
1001+2847	& 3C234	& 0.185	& 1.41 & 113	& $5.33 ^a$	& $5^f$	 \\
1006+3454	& 	& 0.099	& 1.4 & 1896	& $6.67 ^d$	&  \\	
1007+0030	& 	& 0.095	& 1.4 & 17	& $2    ^c$	&  \\	
1016+6014	& 	& 0.031	& 1.4 & 100	& $6    ^c$	&  \\	
1016+4046	& 	& 0.128	& 1.4 & 139	& $2.93 ^d$	&  \\	
1020+4832	& 	& 0.053	& 1.4 & 675	& $1.7  ^d$	&  \\	
1032+5644	& 	& 0.045	& 1.4 & 389	& $6.7  ^c$	&  \\	
1036+0006	& 	& 0.097	& 1.4 & 50	& $4.5  ^c$	&  \\	
1039+0510	& 	& 0.068	& 1.4 & 24	& $3.9  ^c$	&  \\	
1059+0517	& 	& 0.035	& 1.4 & 239	& $3.7  ^c$	&  \\	
1111+4050	& 	& 0.074	& 1.4 & 322	& $1.26 ^d$	&  \\	
1116+2915	& 	& 0.049	& 1.4 & 130	& $2.5  ^d$	&  \\	
1137+6120	& 	& 0.111	& 1.4 & 230	& $4.1  ^d$	&  \\	
1144+3710	& 	& 0.115	& 1.4 & 75	& $2.67 ^d$	&  \\	
1153+0329	& 	& 0.079	& 1.4 & 13	& $2.5  ^c$	&  \\	
1155+5454	& 	& 0.050	& 1.4 & 216	& $5.7  ^d$	&  \\	
1214+0528	& 	& 0.078	& 1.4 & 25	& $5.2  ^c$	&  \\	
1217+0336	& 	& 0.077	& 1.4 & 72	& $4.3  ^c$	&  \\	
1217+0339	& 	& 0.078	& 1.4 & 24	& $1    ^c$	&  \\	
1218+5026	& 	& 0.200	& 1.4 & 210	& $4.4 \pm 0.6^b$	& $50 \pm 15^b$ \\
1252+0315	& 	& 0.099	& 1.4 & 24	& $2.4^c$	&  \\	 
1254+5305	& 	& 0.054	& 1.4 & 33	& $7.3^c$	&  \\	 
1254+2737	& 	& 0.086	& 1.4 & 69	& $1 ^d$	&  \\	 
1302+6229	& 	& 0.076	& 1.4 & 18	& $2.6^c$	&  \\	 
1321+4235	& 3C285	& 0.079	& 1.65 & 184	& $1.2^a$	& $40^f$	 \\ 
1328-0307	& 	& 0.085	& 1.4 & 289	& $6.1^c$	&  \\	 
1330-0206	& 	& 0.087	& 1.4 & 20	& $1.3^c$	&  \\	 
1331-0252	& 	& 0.087	& 1.4 & 13	& $2.5^c$	&  \\	 
1350+2816	& 	& 0.072	& 1.4 & 60	& $2.4 \pm 0.3^b$	& $21 \pm 4^b$ \\
1352+3126	& 3C293	& 0.045	& 1.52 & 256	& $3.86^a$	& $1.9^e$ \\
1354+0528	& 	& 0.077	& 1.4 & 14	& $2.9^c$	&  \\
1400+3021	& 	& 0.206	& 1.4 & 120	& $3.6 \pm 0.8^b$	& $125 \pm 25^b$ \\
1443+5201	& 3C303	& 0.141	& 1.45 & 47	& $1.17^a$	& \\
1454+1620	& 	& 0.046	& 1.4 & 660	& $2^d$	& \\
1504+2600	& 3C310	& 0.054	& 1.45 & 305	& $1.375^a$	& $77.3^e$ \\
1508+5415	& 	& 0.096	& 1.4 & 15	& $3.3^c$	&   \\
1516+0015	& 	& 0.052	& 1.4 & 328	& $3.9^d$	& \\
1524+5428	& 3C319	& 0.192	& 1.65 & 109	& $2.795^a$	& $30^f$ \\
1531+2404	& 3C321	& 0.096	& 1.51 & 307	& $10^a$	&  \\
1552+2005	& 3C326	& 0.089	& 1.40 & 1206	& $3.372^a$	& \\
1557+5440	& 	& 0.047	& 1.4 & 200	& $42.1^c$	&   \\
1602+0158	& 	& 0.104	& 1.4 & 606	& $1^d$	&  \\
1617+3222	& 3C332	& 0.152	& 1.4 & 800	& $4.8 \pm 1.3^a$	& $47 \pm 9^f$ \\
1700+3008	& 	& 0.035	& 1.4 & 135	& $2.2 \pm 0.2^b$	& $48 \pm 9^b$ \\
1713+6402	& 	& 0.079	& 1.4 & 13	& $1.9^c$	&  \\
1728+3146	& 3C357	& 0.166	& 1.4 & 60	& $3 \pm 0.6^b$	& $27 \pm 5^b$ \\
1729+5415	& 	& 0.084	& 1.4 & 28	& $3^c$	&  \\
2153-0711	& 	& 0.059	& 1.4 & 183	& $3.6^c$	&  \\
2157-0750	& 	& 0.062	& 1.4 & 28	& $2^c$	&  \\
2214+1350	& 3C442A & 0.026 & 1.38 & 597	& $2^a$	&  \\
2231-0824	& 	& 0.083	& 1.4 & 111	& $2.5^c$	&  \\
2315-0026	& 	& 0.091	& 1.4 & 25	& $4.6^c$	&  \\
2333-0027	& 	& 0.059	& 1.4 & 81	& $8^c$	&  \\
\hline
%\end{tabular}
\caption{FR-II radio sources. Sizes and morphologies come from: (a) Leahy, Bridle \& Strom 2000; (b) Machalski \etal\/ 2004; (c) Shabala \etal\/ 2008; (d) Gendre \& Wall 2009. Complementary radio source ages, where available, are from: (e) Alexander \& Leahy 1987; (f) Rawlings \& Saunders 1991; (g) Orr\`u \etal\/ 2010. Sources 1217+0336 and 1217+0339 are unrelated nearby radio sources.}
\label{tab:FR2s}
\end{longtable}

%\begin{table*}
%\centering
%\begin{tabular}{lcccc}
%\hline
%\multicolumn{1}{l}{IAU} & Other & redshift & LAS & $t_{\rm syn}$ \\ 
%\multicolumn{1}{l}{name} & name & & (arcsec) & (Myr) \\
%\hline
\begin{longtable}{llllll}
\hline
IAU & Other & redshift & frequency & LAS & $t_{\rm syn}$ \\ 
name & name & & (GHz) & (arcsec) & (Myr) \\
\hline
\endhead
\multicolumn{2}{l}{}\\
\hline
IAU & Other & redshift & frequency & LAS & $t_{\rm syn}$ \\ 
name & name & & (GHz) & (arcsec) & (Myr) \\
\hline
\endfirsthead
\multicolumn{2}{l}{{\it continued on next page}}
\endfoot
\multicolumn{2}{l}{}
\endlastfoot
0318+4151 &	NGC 1265 & 0.025 & 1.38 & $1216^a$	&\\
0319+4130 &	3C84	 & 0.017 & 1.38 & $1350^a$	&\\
0847+5352 &		 & 0.045	& 1.4 & $103^d$	&\\
0901+2901 &		 & 0.194	& 1.4 & $50^d$	&\\
1002+1951 &		 & 0.168	& 1.4 & $74^d$	&\\
1031+5225 &		 & 0.166	& 1.4 & $89^d$	&\\
1140+1203 &		 & 0.081	& 1.4 & $109^d$	&\\
1145+1936 &	3C264	 & 0.022	& 1.40 & $522^a$	& $10^f$\\
1225+1253 &	3C272.1	 & 0.005	& 1.41 & $189^a$	&\\
1229+1144 &	A1552	 & 0.086	& 1.49 & $168^a$	&\\
1230+1223 &	3C274	 & 0.005	& 1.46 & $836^a$	&\\
1236+1632 &		 & 0.068	& 1.4 & $208^d$	&\\
1316+0702 &		 & 0.051	& 1.4 & $840^d$	&\\
1342+0504 &		 & 0.136	& 1.4 & $89^d$	&\\
1416+1048 &	3C296	 & 0.025	& 1.45 & $347^a$	&\\
1449+6316 &	3C305	 & 0.042	& 1.42 & $14^a$	&\\
1504+2835 &		 & 0.043	& 1.4 & $502^d$	&\\
1513+2607 &	3C315	 & 0.108	& 1.42 & $201^a$       & $53.4^e$\\
1516+0701 &		 & 0.034	& 1.4 & $89^d$	&\\
1617+3500 &	NGC 6109 & 0.030	& 0.609 & $890^a$	&\\
1628+3933 &	3C338	 & 0.030	& 4.89 & $117^a$	& $30^f$\\
\hline
%\end{tabular}
\caption{FR-I radio sources. References are as in Table~\ref{tab:FR2s}.}
\label{tab:FR1s}
\end{longtable}

\subsection{Galaxy selection}
\label{sec:separateSamples}

\subsubsection{Photometric redshifts}

\begin{figure}
\centering
  \includegraphics[height=0.45\textwidth,angle=0]{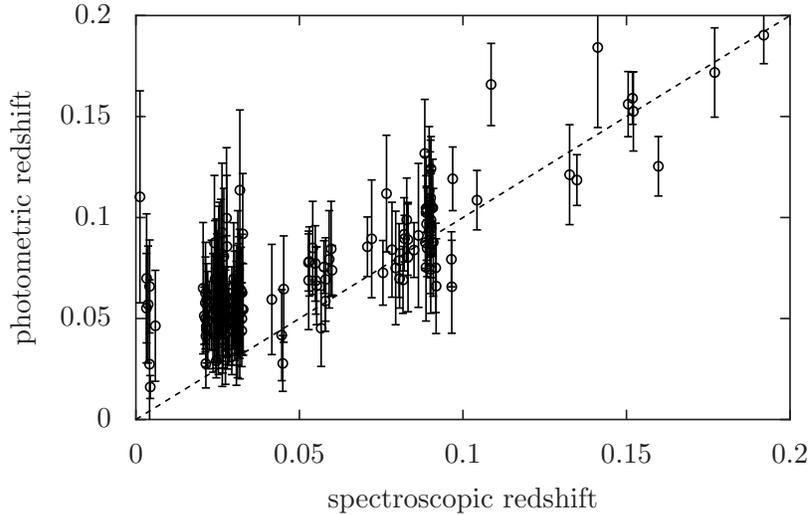}
\caption{Comparison of spectroscopic and photometric redshifts for galaxies near 33 radio sources. The measurements agree within photometric redshift uncertainties for the bulk of the sample at $z \geq 0.05$, but diverge at lower redshift.}
\label{fig:photoz}
\end{figure}

Galaxies located near radio sources were selected from SDSS. In this paper we are interested in the effects of radio sources on photometric properties of galaxies, and a photometric (rather than spectroscopic) sample was therefore employed. While this greatly increases sample statistics, care must be taken in assessing the accuracy of photometric redshift estimates. In Figure~\ref{fig:photoz} we plot spectroscopic and photometric redshifts for a subsample of 161 galaxies, derived from the SDSS pipeline. The photometric redshifts are estimated using template fitting. These spectral energy distribution templates are obtained using galaxies with spectroscopic identifications, and extended with spectral synthesis models \cite{AdelmanMcCarthy07}. Photometric redshifts are in good agreement with spectroscopic values for $z \geq 0.05$, with most values agreeing within the uncertainties of photometric measurements. This agreement disappears at low redshifts. In our sample, 12 of 21 FR-I radio sources have $z<0.05$, compared to only 11 of 72 FR-IIs. This is a known issue with the SDSS pipeline \cite{AdelmanMcCarthy07}, and is related to the inherent difficulty in getting spectroscopic redshifts to be more accurate than $\Delta z \sim 0.02$. It is therefore important to consider whether any systematic effects are introduced into the analysis by adopting photometric redshifts. Specifically, there are three crucial considerations: completeness of the sample, contamination, and selection biases. 

\vspace{0.1cm}

{\it Completeness}

An estimate of how many bona fide group members are not picked up by our photometric identification procedure comes from comparing galaxy counts obtained using spectroscopic and photometric redshifts. We selected all spectroscopically identified galaxies projected to be in or near the 26 largest radio sources in our sample. The true number of galaxies associated with the group was estimated by requiring that the spectroscopic redshifts of group members were not offset from the redshift of the radio source host by more than $\Delta z = 0.02$. This $\Delta z$ is the optimal tolerance value: relaxing the redshift restriction (by increasing $\Delta z$) in an individual group does not change the number counts appreciably; however, decreasing $\Delta z$ gives an appreciable decrease in the number counts, indicating that some group members are being excluded. The number of photometric matches was then obtained by requiring that the spectroscopic redshift of each galaxy lay within the photometric $1\sigma$ uncertainties. 

\begin{figure}
\centering
  \includegraphics[height=0.45\textwidth,angle=0]{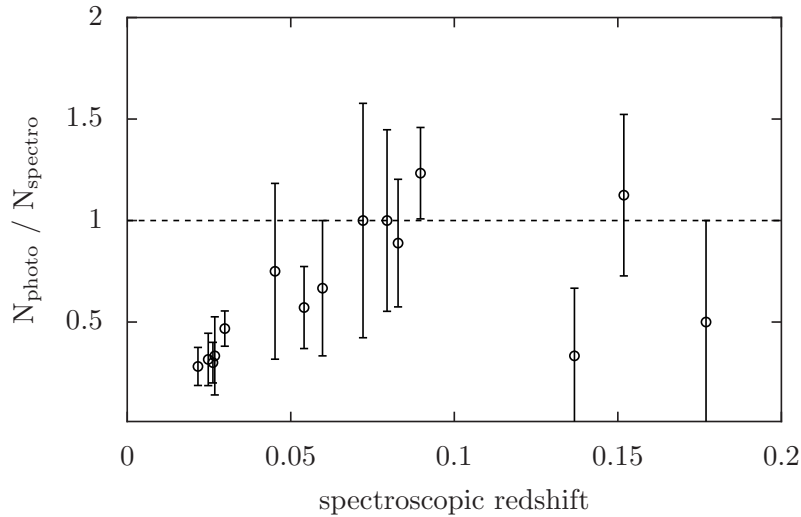}
\caption{Photometric sample completeness for galaxies near large radio sources. Spectroscopic and photometric methods select galaxies in a similar way at $z \geq 0.04$. The photometric samples can be significantly incomplete (as low at 30\%) at lower redshift. Only groups with more than one spectroscopic identification are shown.}
\label{fig:spectroPhotoRatio}
\end{figure}

Figure~\ref{fig:spectroPhotoRatio} shows the redshift dependence of the photometric-to-spectroscopic matching rates. At $z \geq 0.04$ the completeness is close to unity. At $z<0.04$ the completeness can be as low as 30 percent.

\vspace{0.1cm}

{\it Contamination}

Incorrect redshifts can also result in inclusion of unrelated galaxies seen in projection. The largest radio source in our sample, 3C\,236, is 32 arcmin in size. At a redshift of 0.1, and with a redshift uncertainty of 0.02, this corresponds to a co-moving volume of 960 Mpc$^3$. For comparison, the local number density of groups has been estimated by Yang \etal\/ \shortcite{YangEA05} to be around $10^{-3}$~Mpc$^{-3}$. A similar number is obtained from simulations for the number density of local dark matter haloes with masses in excess of $10^{13}$~\Msun. Most of these groups contain only a single member, with only 3 percent harbouring four or more galaxies \cite{YangEA05}, and thus the expected contamination is of order one to a few galaxies. Since most groups subtend a much smaller solid angle on the sky than 3C\,236, contamination is not expected to be a major issue in the present work. 

\vspace{0.1cm}

{\it Selection effects}

It is important to examine whether adopting photometric redshifts preferentially selects against galaxies with certain properties. Galaxy properties are often a strong function of mass, and galaxy colours are of particular interest in the present work. We therefore investigated whether the offset between photometric and spectroscopic redshift values correlates with either colour or $r$-band magnitude (a proxy for stellar mass; e.g. Shabala \etal\/ 2008). We found no such correlations, suggesting that despite the lack of completeness at low redshift, photometric redshift identifications are suitable for our analysis.

It is further worth noting that because in our analysis below we compare similar galaxies inside and outside the radio contours, any major redshift-dependent systematic effects are unlikely. In the following we therefore include all galaxies for which the redshift of the target radio source lies within the uncertainty of the galaxy photometric redshift.

\subsubsection{Classification}

\begin{figure}
\centering
  \subfigure[FR-II]{\includegraphics[height=0.35\textwidth,angle=0]{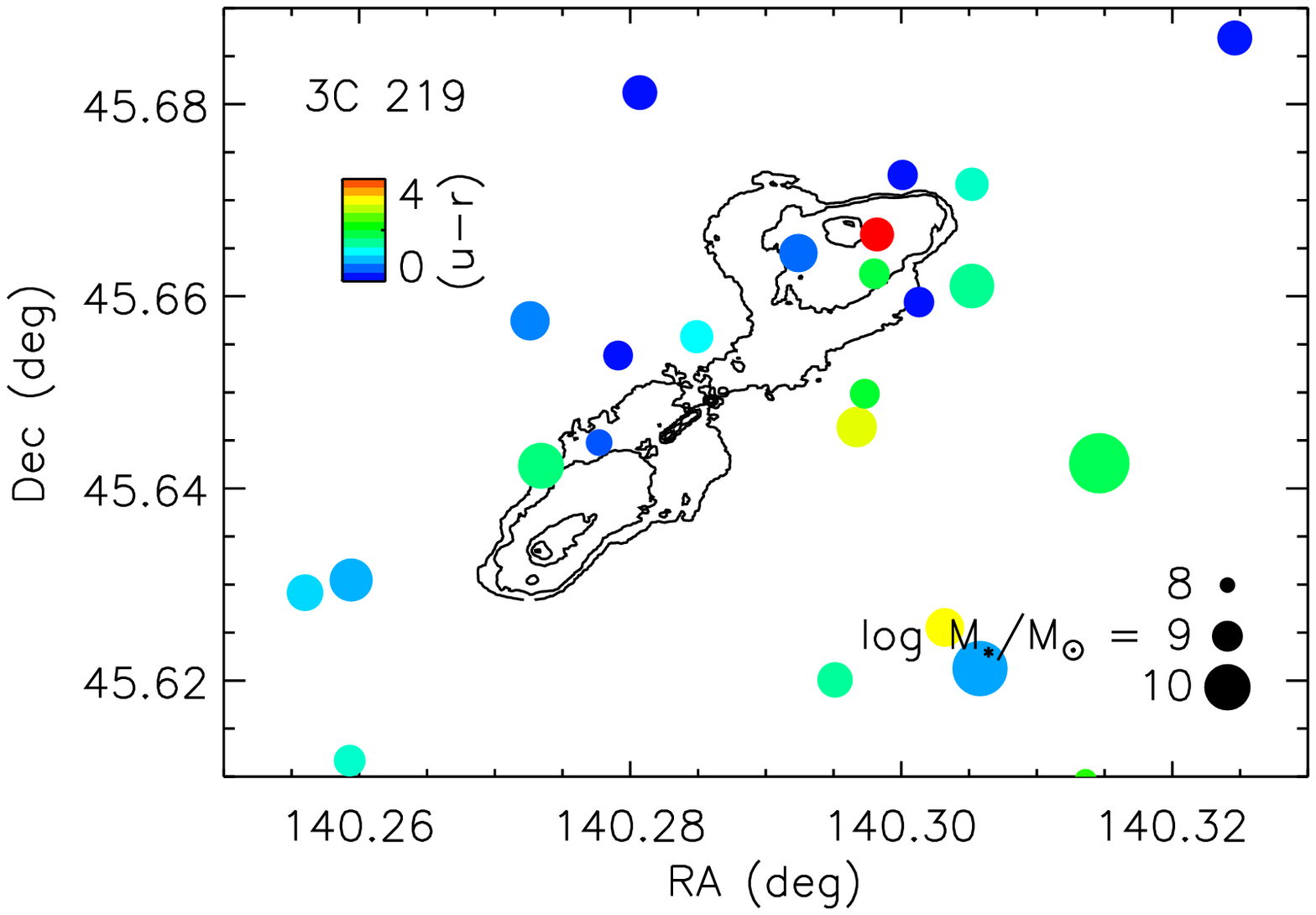}}
  \subfigure[FR-I]{\includegraphics[height=0.35\textwidth,angle=0]{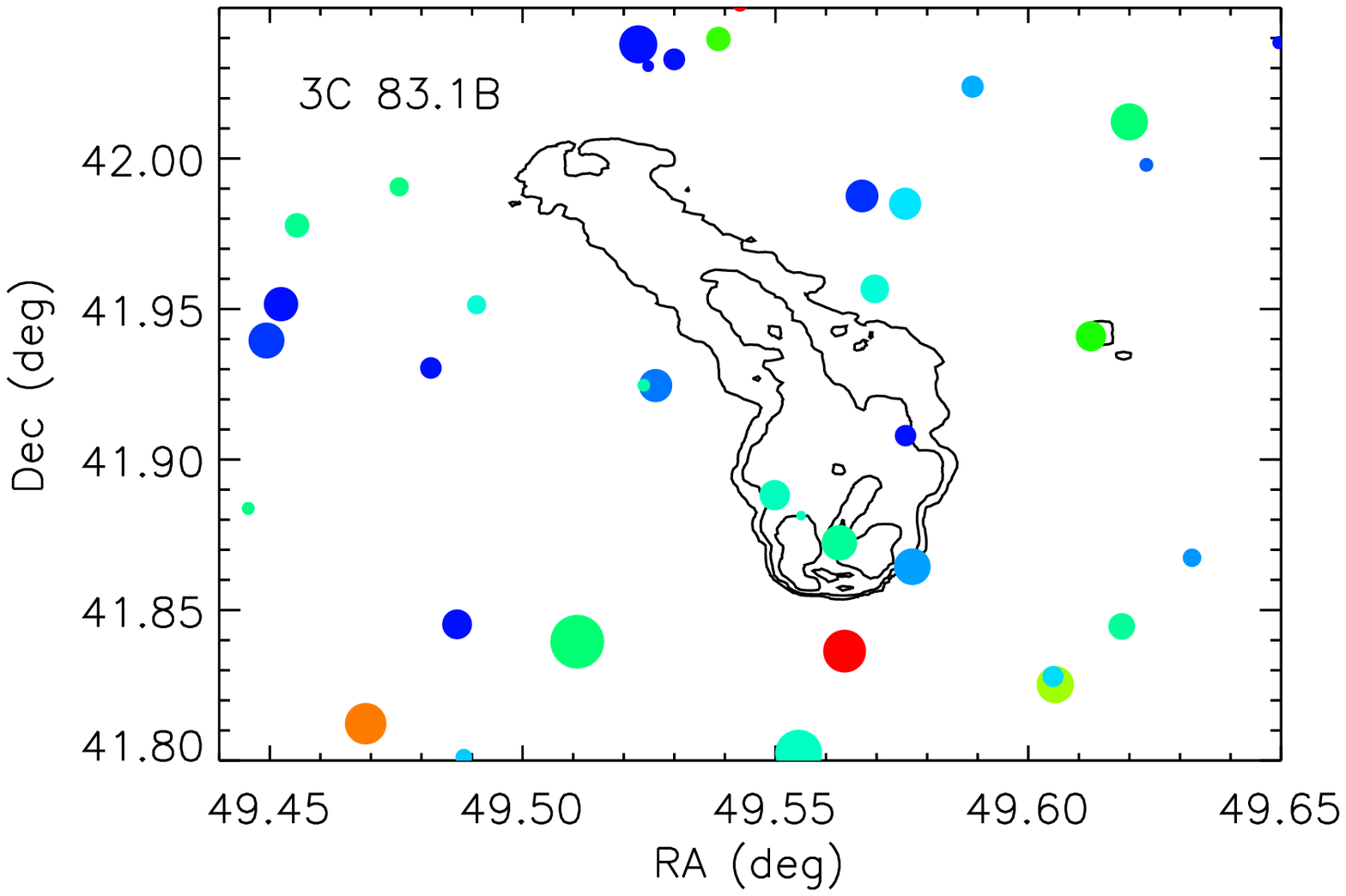}}
\caption{Example of galaxy colours around FR-II (3C219) and FR-I (3C83.1B) radio sources. Contours correspond to 1, 3, 10, 30, 90\% of maximum emission, and are available at www.jb.man.ac.uk/atlas. VLA 1.6 GHz observations for 3C\,219 are taken from Clarke et al. (1992); and the VLA 1.4 GHz observations of 3C83.1B are from O'Dea (unpublished). For each galaxy, circle size corresponds to galaxy mass, while $(u-r)$ colour is represented using a colour table.}
\label{fig:contoursExample}
\end{figure}

SDSS photometry was obtained for objects within a radius equal to 1.5 times the maximum angular extent of the radio source (column 5 in Tables~\ref{tab:FR2s} and \ref{tab:FR1s}) from the centre of the radio source. Galaxies with photometric redshifts matching the spectroscopic redshift of the radio source host (i.e. with $z_{\rm gal}-\sigma_{z_{\rm gal}} \leq z_{\rm radio} \leq z_{\rm gal}+\sigma_{z_{\rm gal}}$) were retained for analysis. Radio maps were used to detemine whether individual galaxies are (in projection) located inside or outside the radio contours. Where available, high resolution radio source maps at 1.4-1.6 GHz (Leahy, Bridle \& Strom 2000) were used. FIRST/NVSS maps at 1.4 GHz were adopted for the rest of the sources, with the exception of FR-I sources 3C\,338 (4.89 GHz map used) and NGC\,6109 (608 MHz). For these two objects, the lower resolution FIRST maps at 1.4 GHz show similar source size and structure to the above maps, and the higher resolution versions were therefore used. Examples are shown in Figure~\ref{fig:contoursExample}.

\vspace{0.1cm}

{\it{FR-II sources}}

Galaxies lying within the radio source contours are subject to projection effects due to the uncertainty in radio source orientation. In other words, these galaxies could lie either immediately behind or in front of the radio emission, but appear to be coincident with this emission. This problem can be mitigated for the FR-II (classical double) radio sources. These objects typically have cylindrical morphologies, at least in regions away from the hotspots. The aspect ratio $R_{\rm T}$ between semi-major and semi-minor axes is given for each source in column 5 of Table~\ref{tab:FR2s}. For $D_{\rm max}$ being the maximum linear extent of the FR-II source, galaxies lying within $R_{\rm def} = D_{\rm max} / 2 R_{\rm T}$ of the radio cource centre will be immune from such a projection effect, since they fall within the radio cocoon regardless of the projection angle.

Projection effects due to uncertainty in galaxy position in redshift space are much harder to quantify. However, as most radio AGN reside in cluster centres and are quite extended in the direction perpendicular to the jet (i.e. $R_{\rm def}$ is comparable with the cluster core radius), we expect the majority of the galaxies that appear to lie within $R_{\rm def}$ of the cluster centre to do so.

Motivated by these considerations, we separated galaxies associated with FR-II radio sources into three categories. Galaxies lying outside the radio emission (in projection) were allocated to the ``outside'' group. Galaxies projected to lie within the radio source contours but outside $R_{\rm def}$ of the cluster centre were placed in the ``mixed'' group. Some of these would presumably be physically coincident with the radio emission, while others won't. Finally, galaxies located within $R_{\rm def}$ of the cluster centre were placed in the ``inside'' group. These correspond to objects that are very likely to have been overrun by the radio source- only galaxies located well away from the cluster core but seen close to the line of sight to the core can contaminate this sample.

\vspace{0.1cm}

{\it{FR-I sources}}

It is much more difficult to perform a similar analysis for the edge-darkened FR-I sources. These primarily consist of two types of objects which are often referred to in the literature as twin-jet radio AGN and relaxed doubles.

The former initially appear as two jets emanating from the central engine. The jets are eventually disrupted (usually at a well-defined flare point), and at larger radii the radio emission turns into a plume due to interaction with the intra-cluster medium. 3C296 is a classical example of this type of source. We include narrow and wide-angled tail galaxies (e.g. 3C264) in this subclass of FR-Is, since these are most likely twin-jet radio sources that have been swept up by interaction with the ICM through which they are travelling.

Relaxed doubles are sources in which radio emission declines gradually with radius, in a quasi-radially symmetric fashion. These objects do not exhibit any compact structure, with 3C84 being a famous example.

We identify $R_{\rm def}$ with the flare point for twin-jet sources, and with half the maximal radial extent (i.e. $R_{\rm def}=D_{\rm max}/4$) for relaxed doubles. The galaxies associated with FR-I radio sources are then split into ``outside'', ``inside'' and ``mixed'' regions in exactly the same way as those near FR-IIs. As we show below, FR-Is do not appear to exhibit any difference in properties between the three classes, suggesting our classification is adequate.

\section{Results}
\label{sec:results}

The major purpose of this work is to study the effects of radio sources on photometric properties of galaxies, and in particular galaxy colours. We therefore stacked all galaxies into the three groups outlined above. Around FR-II radio sources this yielded 58 galaxies in the ``inside'' category, 6678 in the ``outside'' group, and 318 in the ``mixed''. For FR-Is, 27 galaxies were found in the ``inside'' group, 2149 in the ``outside'', and 36 in ``mixed''.

\subsection{Galaxy colours}
\label{sec:colours}

Galaxy properties such as colour are a strong function of mass, with massive galaxies preferentially being redder. We therefore restricted our analysis to objects within a well-defined mass range $10^8 \leq M_\star \leq 10^{10}$~\Msun. Stellar mass was approximated via the $r$-band absolute magnitude, assuming that the solar bolometric correction is applicable (see Shabala \etal\/ 2008 for details).

\begin{figure}
\centering
  \includegraphics[height=0.45\textwidth,angle=0]{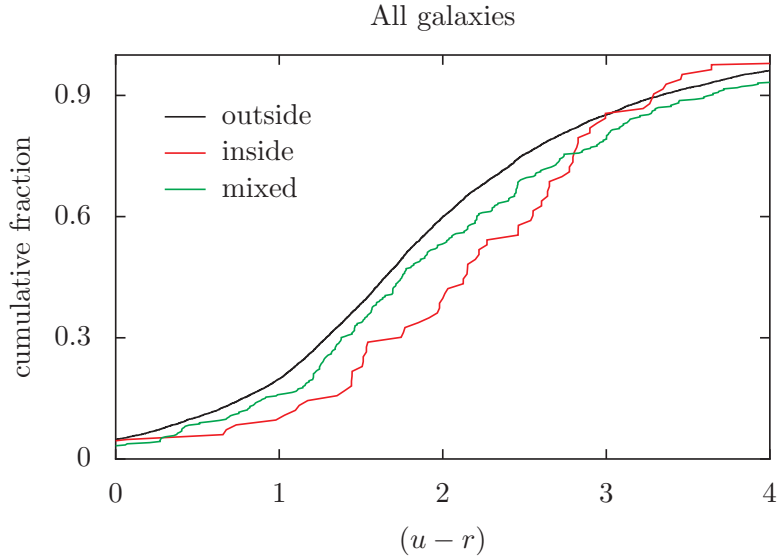}
\caption{Galaxy $(u-r)$ colours, stacked for all radio sources. The three regions are defined in the text. Galaxies within the path of the radio source show a clear offset to redder colours.}
\label{fig:coloursAll}
\end{figure}

In Figure~\ref{fig:coloursAll} we plot the cumulative distribution of $(u-r)$ $k$-corrected colours for the three groups, stacking galaxies around all radio sources in our sample. There is a clear offset between the ``inside'' and ``outside'' groups, with the ``inside'' galaxies exhibiting redder colours.

\begin{figure}
\centering
  \subfigure[FR-II]{\includegraphics[height=0.35\textwidth,angle=0]{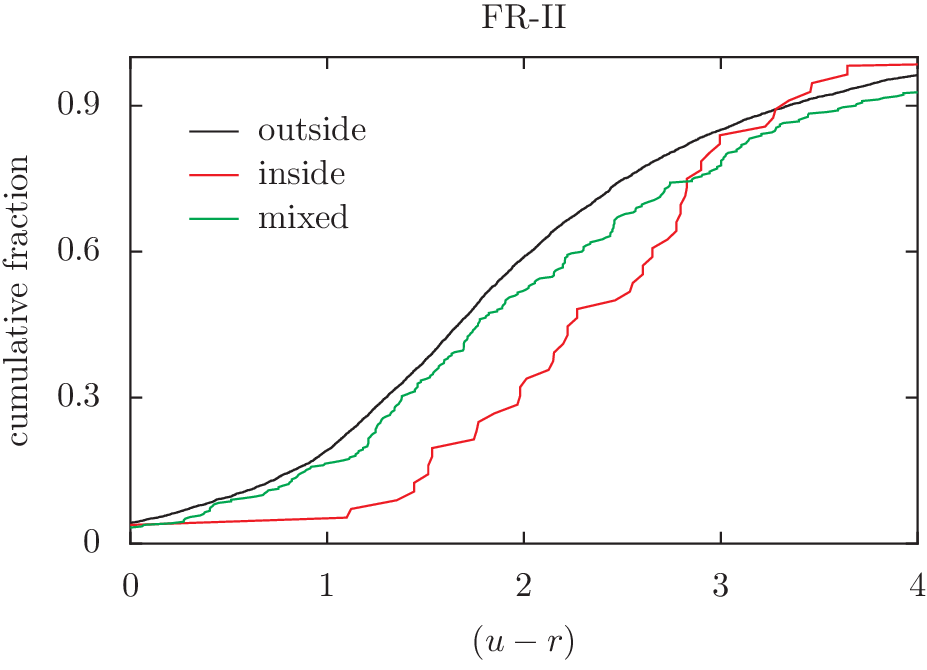}}
  \subfigure[FR-I]{\includegraphics[height=0.35\textwidth,angle=0]{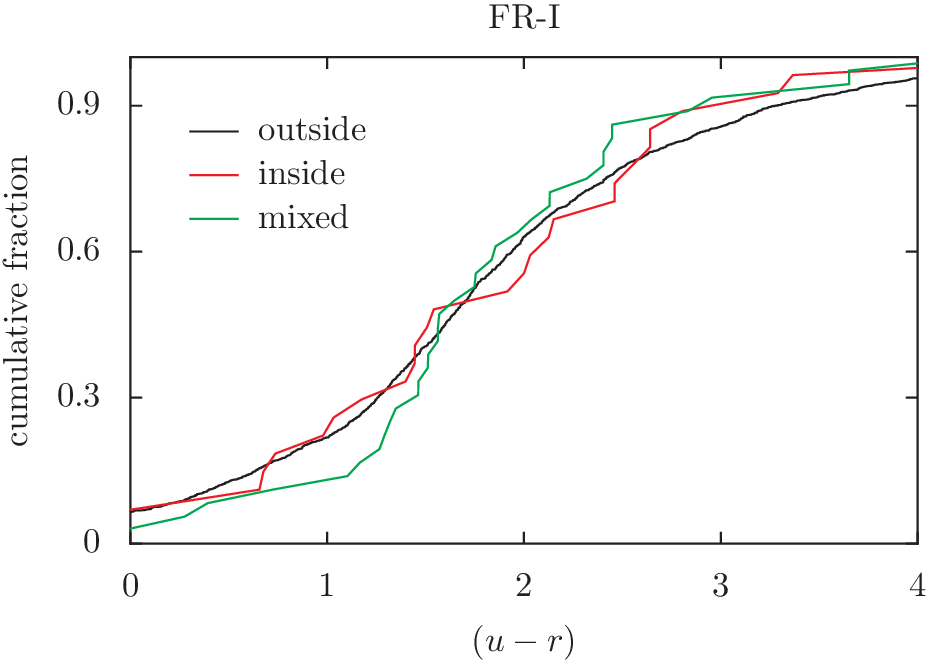}}
\caption{Same as Figure~\ref{fig:coloursAll} but separating the sample into FR-I and FR-II radio sources. The statistically significant offset to redder colours for galaxies overrun by radio sources remains for FR-IIs, but disappears for FR-Is.}
\label{fig:coloursFRtypes}
\end{figure}

Splitting up the sample into FR-I and FR-II radio sources (Figure~\ref{fig:coloursFRtypes}), it can be seen that this colour difference is driven exclusively by the FR-II population. In other words, FR-IIs appear to affect their environment, while FR-Is do not. These conclusions are confirmed by Komogorov-Smirnov (KS) tests, with the observed FR-II offset being statistically significant at the 2.5\% level, and no statistically significant differences found for FR-Is.

\subsection{Mass distribution}
\label{sec:massDist}

\begin{figure}
\centering
  \subfigure[FR-II]{\includegraphics[height=0.349\textwidth,angle=0]{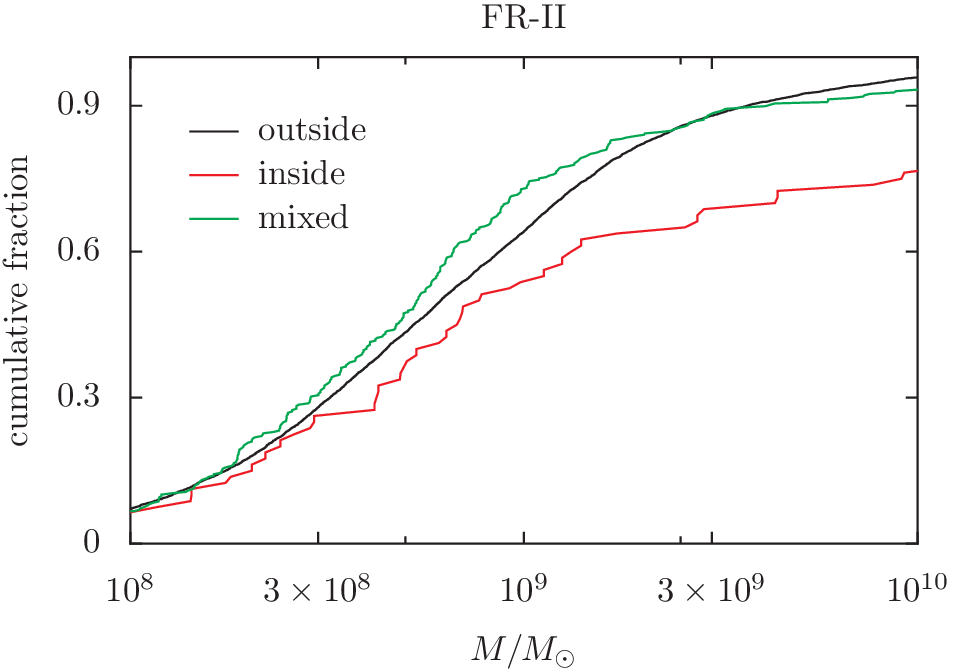}}
  \subfigure[FR-I]{\includegraphics[height=0.349\textwidth,angle=0]{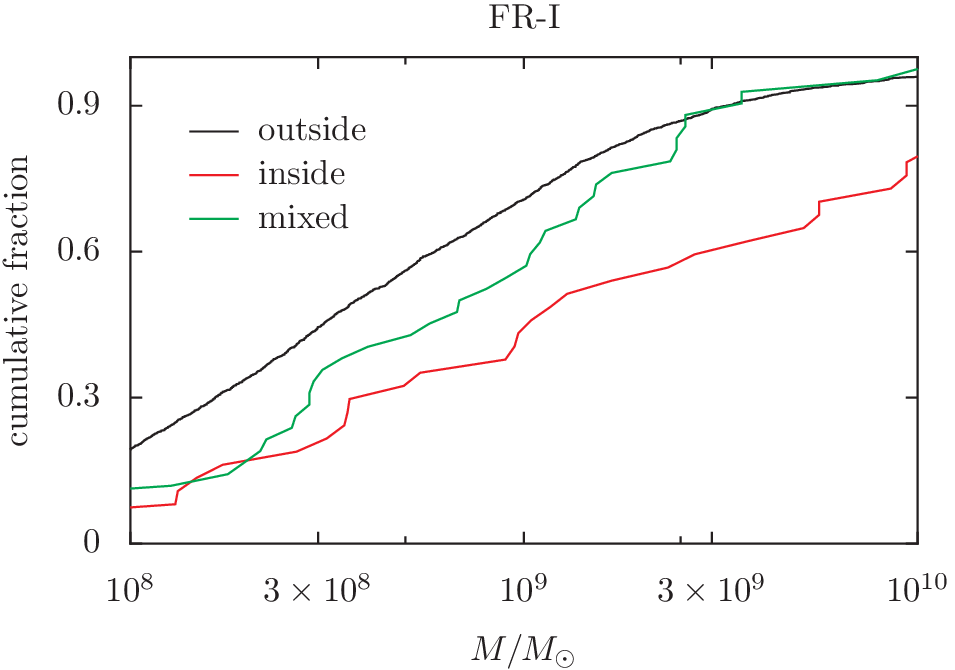}}
\caption{Mass distribution for the FR-II (left panel) and FR-I (right) samples. The ``inside'' group is biased towards more massive galaxies.}
\label{fig:massDist}
\end{figure}

Care must be taken in interpreting these results. All radio sources in our sample are located at the centres of groups or clusters. Mass segregation means that galaxies near the cluster centres (and therefore more likely to have been classified in the ``inside'' category) will be more massive, and therefore redder. Figure~\ref{fig:massDist} shows that this is indeed the case for both the FR-I and FR-II samples.

\begin{figure}
\centering
  \subfigure[FR-II]{\includegraphics[height=0.35\textwidth,angle=0]{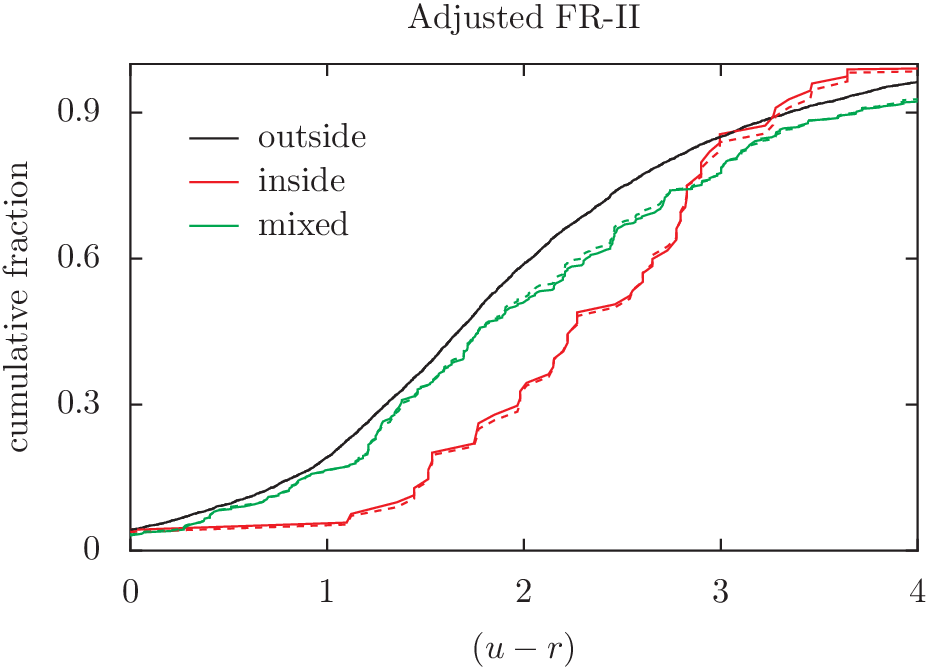}}
  \subfigure[FR-I]{\includegraphics[height=0.35\textwidth,angle=0]{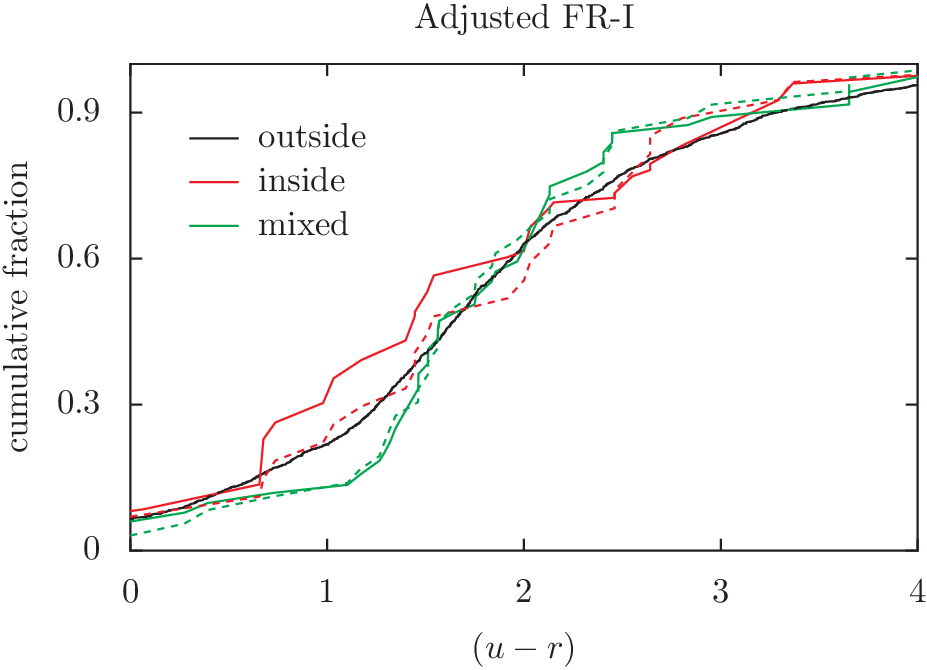}}
\caption{$(u-r)$ colour distributions for galaxies around radio sources. The colours are corrected for mass distribution in the range $10^8 \leq M_\star \leq 10^{10}$~\Msun. Solid lines correspond to corrected distributions; dashed lines of the same colour to uncorrected ones. The corrections are negligible for FR-IIs, and the colour offset remains statistically significant for this subsample. For FR-Is the corrections are more significant due to smaller number statistics, however as in Figure~\ref{fig:coloursFRtypes} there is still no difference in colour distribution between the three groups.}
\label{fig:coloursFRtypesCorr}
\end{figure}

We correct for this effect in two ways. Firstly, our analysis is restricted to galaxies with $M_\star < 10^{10}$~\Msun. In other words, the most massive, red galaxies that would drive redder colours for the ``inside'' category are excluded. Furthermore, we break up the $10^8 \leq M_\star \leq 10^{10}$~\Msun\/ mass range into bins and weigh the colours by the ratio of number counts between the category of interest (``inside'' or ``mixed'') and the control sample (``outside''). To put it another way, we mimic the mass profile of the ``outside'' category. Figure~\ref{fig:coloursFRtypesCorr}, confirmed by KS tests, shows that our findings are robust to these corrections.

\section{Discussion}
\label{sec:discussion}

The results of Section~\ref{sec:results} appear to suggest that FR-II and FR-I radio sources interact with their environments in entirely different ways. A large body of observational evidence suggests that FR-II sources grow by driving large-scale bow-shocks (e.g. Fabian \etal\/ 2003, Croston \etal\/ 2009), sweeping up the ambient gas as they proceed. Shock heating and gas uplifting away from the centre of the potential well are seen in analytical models and numerical simulations \cite{KA97,HRB98,BassonAlexander03}, and are in fact necessary to solve the cooling flow problem \cite{BinneyTabor95} and reconcile observations of the local stellar mass function with galaxy formation models \cite{SA09b}. While the impact of powerful (i.e. FR-II) radio source feedback on cluster gas is well understood, it is not immediately obvious that such feedback can affect individual galaxies in a similar way. A propagating bow shock will compress and shock heat both the atomic and molecular gas reservoirs that it overruns. If the local gas density is above some critical value, the gas clumps will become radiative (e.g. Sutherland \& Bicknell 2007) and the shock can in fact {\it enhance} star formation, rather than suppress it \cite{AntonuccioDeloguSilk08}. It is a different story if one is concerned with diffuse H\,I gas reservoir, however, which is more susceptible to both shock heating (due to longer cooling times) and uplifting.

FR-I radio sources start their lives as FR-IIs, but have their jets disrupted by interaction with the dense interstellar/intergalactic medium. As a result, the bow shocks typically do not propagate far outside the host galaxy, and it is difficult for FR-Is to affect their environment other than by turbulent mixing of the radio plasma with the cluster gas. The findings of the previous section strongly suggest that it is the propagation of the bow shock that is responsible for the redder colours of galaxies. The exact feedback mechanism (i.e. heating or gas expulsion) is not important to the present discussion. Either way, FR-IIs provide a way of suddenly depleting a reservoir of gas previously available for star formation.

\subsection{Feedback model}
\label{sec:model}

\subsubsection{Star formation and colours}
\label{sec:modelSF}

We set out to test whether gas heating/removal associated with an FR-II radio source can explain the observed distribution of galaxy colours. The first step is to construct a star formation history for a galaxy unaffected by AGN feedback. Our modelling follows the prescriptions of Kaviraj \etal\/ \shortcite{KavirajEA07}, which have proved very successful at describing the colour evolution of the early-type galaxy population. The star formation histories of galaxies are represented by an instantaneous burst of star formation at high redshift, followed by more gradual recent star formation. The old burst, in which most of the stars are formed, is characterized by its age $t_1$, typically $5-10$ Gyrs (e.g. Kaviraj 2009). Recent star formation is parametrized by the age of the burst $t_2$, the timescale associated with star formation $\tau_{\rm dyn}$ and the initial gas fraction $f_{\rm gas,0}$ available for recent star formation at time $t_2$.

The star formation rate is given by the Schmidt-Kennicutt law,

\begin{equation}
  \psi = (\epsilon/\tau_{\rm dyn}) M_{\rm gas}
\label{eqn:SK_sfr}
\end{equation}
with star formation efficiency $\epsilon = 0.02$ and dynamical timescale $\tau_{\rm dyn} = 0.05$~Gyrs. Apart from being physically sensible, Kaviraj \etal\/ \shortcite{KavirajEA11} find that these values reproduce the observed number counts of blue cloud and red sequence galaxies. We note that, being an empirical relation, the Schmidt-Kennicutt law implicitly includes the effects of supernova feedback on star formation. Equating the star formation rate with the rate of change of gas mass yields

\begin{equation}
  \frac{\psi}{M_\star} = \frac{\epsilon}{\tau_{\rm dyn}} f_{\rm gas,0} e^{-\epsilon \left( \frac{t}{\tau_{\rm dyn}} \right)}
\label{eqn:fracSFR_noFeedback}
\end{equation}
where $f_{\rm gas,0}$ is the initial gas fraction in the galaxy at time $t_2$.

\begin{figure}
\centering
  \includegraphics[height=0.45\textwidth,angle=0]{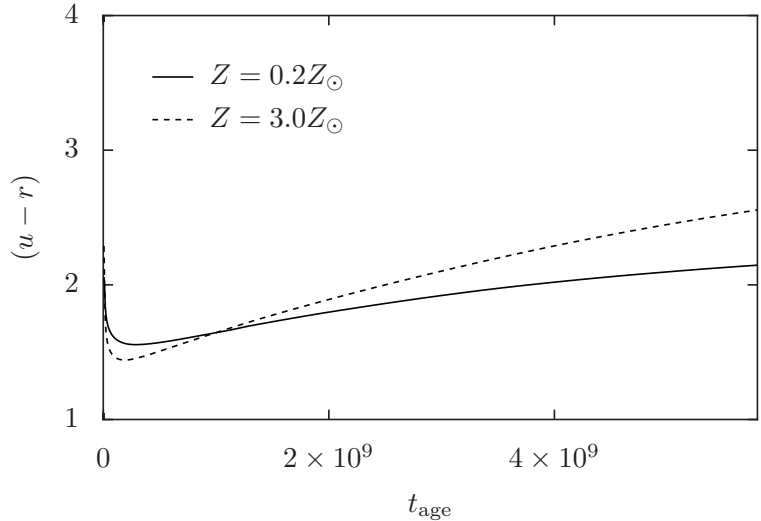}
\caption{$(u-r)$ colour evolution for a star-forming galaxy with an underlying 9 Gyr old population, and recent star formation described by the Schmidt-Kennicutt law (Equation~\ref{eqn:SK_sfr}) with $t_2=2.8$~Gyrs and $f_{\rm gas,0}=0.07$. Solid lines are for metallicity of 0.2~Z$_\odot$, and dashed for 1~Z$_\odot$.}
\label{fig:exampleColourTracks}
\end{figure}

Stellar population synthesis models can then be used to infer the colour evolution of the galaxy. We adopt the Bruzual \& Charlot \shortcite{BruzualCharlot03} models with a Salpeter Initial Mass Function (IMF). Galaxy colours are computed from luminosities, adding up contributions from the old underlying stellar population and the more recently formed stars. Figure~\ref{fig:exampleColourTracks} shows example tracks for $(u-r)$ colour evolution with $t_1=9$~Gyrs, $t_2=6$~Gyrs, $f_{\rm gas,0}=0.07$ and metallicities of $Z=0.2$ and $1 Z_\odot$. 

Comparison with Figure~\ref{fig:coloursFRtypes} shows that the model and observations span a similar range in colour. In reality, the parameters $t_1$, $t_2$, $f_{\rm gas}$ and $Z$ will differ from galaxy to galaxy. We follow the study of local LIRGs by Kaviraj \shortcite{Kaviraj09} in adopting distributions for $t_1$ and $t_2$. The probability distribution in the old burst timescale $t_1$ spans the range $4.8-8.8$~Gyrs, rising linearly from these values towards a peak at $6.8$~Gyrs. The timescale $t_2$ associated with the onset of recent star formation is modelled with a flat prior to an age of $2.8$~Gyrs. While it is true that these distributions found for LIRGs may not apply to our more general galaxy population, we believe that we can be justified in adopting these. The $(u-r)$ colours are primarily driven by the $u$-band luminosity evolution of the young stellar population, and are therefore not very sensitive to the old burst age. As Figure~\ref{fig:exampleColourTracks} shows, the colour evolution is relatively slow for $t_2>1$~Gyr, and we therefore do not expect the exact form of the prior on $t_2$ to affect our results either. Importantly, while we adopt a log-normal\footnote{As Sutherland \& Bicknell \shortcite{SutherlandBicknell07} note, a log-normal distribution is the limiting distribution for a product of random increments, analogous to the normal distribution playing that role for random additive increments. In our case, such increments could represent various merging and feedback processes that affect the gas fraction.} form for the distribution of the initial gas graction $f_{\rm gas,0}$, the normalization is left as a free parameter used to match the observed colours in unaffected (i.e. ``outside'') galaxies in our sample.

\begin{figure}
\centering
  \includegraphics[height=0.45\textwidth,angle=0]{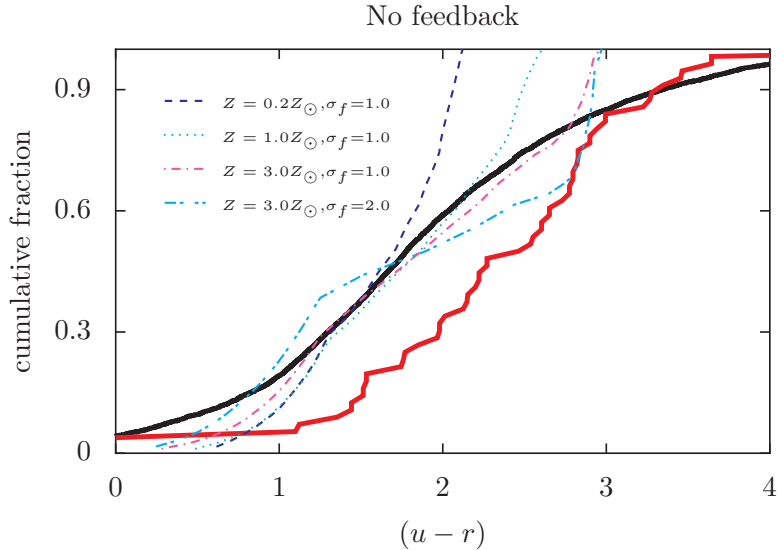}
\caption{$(u-r)$ model colour distributions without AGN feedback. Thick black solid line corresponds to observations of galaxies lying outside the FR-II radio sources. Thick red solid line shows observations of galaxies that have been overrun by FR-II radio sources. Dashed and dotted lines denote various models, with mean initial gas fraction fixed at $<f_{\rm gas,0}>=0.07$ but varying scatter about this value and metallicity. While a good fit can be obtained for the bulk of the ``outside'' galaxy population, no reasonable gas fraction distribution can reproduce the colours of galaxies affected by FR-II feedback.}
\label{fig:outsideFit}
\end{figure}

Figure~\ref{fig:outsideFit} shows the best-fit models to $(u-r)$ colours for the ``outside'' galaxies near FR-II radio AGN. A range of metallicities is considered. Reasonable gas fractions of $<f_{\rm gas,0}> = 0.07$ and $\sigma_{\rm log\, f_{\rm gas,0}} = 1.0$ reproduce the bulk of the observations. These values are consistent with the observed gas fraction $ \sim 0.1$ in local early-type galaxies \cite{Kaviraj09}. The very blue and very red galaxies are not very well reproduced by the models, most likely due to a number of contributing factors. It is possible that we have not encompassed the full range of metallicities or old starburst ages. As Figure~\ref{fig:outsideFit} shows, up to a 0.5 dex offset in $(u-r)$ colour can arise due to a change in metallicity. Alternatively, tidal stripping effects not considered here may be important: accreting galaxies would exhibit enhanced star formation, while the stripped companions would appear redder. An intriguing possibility for the origin of the reddest galaxies, discussed in Section~\ref{sec:feedbackModes}, is that these carry an imprint of previous AGN outbursts. We note that, overall, the models fit the bulk of the sample well.

Modifying the scatter changes the overall slope as well as shape in the middle of the distribution, but does not systematically shift all galaxies to either redder or bluer colours. The redder colours of ``inside'' galaxies in our sample can only be achieved with very low gas fractions, a requirement that is both at odds with the observed gas fraction values in the local volume \cite{Kaviraj09} and the gas fractions required to explain the colours of unaffected galaxies (lying in the ``outside'' group) in the same clusters.

\subsubsection{AGN feedback}
\label{sec:modelAGN}

In our simple model we assume that the effect of a poweful AGN (such as an FR-II) is to starve the galaxy of gas previously available for star formation. The galaxy star formation history is therefore identical to that described above, except for a sudden truncation a time $t_{\rm AGN}$ ago, where $t_{\rm AGN}$ is the time since an AGN-driven shock has overrun the galaxy. Since typical lobe expansion speeds for classical double radio sources are of the order of a few percent of the speed of light \cite{AlexanderLeahy87}, and most galaxies that are definitely affected by the AGN lie relatively close to cluster centre by construction (see Section~\ref{sec:separateSamples}), $t_{\rm AGN}$ approximately corresponds to the current age of the radio source.

\begin{figure}
\centering
  \includegraphics[height=0.45\textwidth,angle=0]{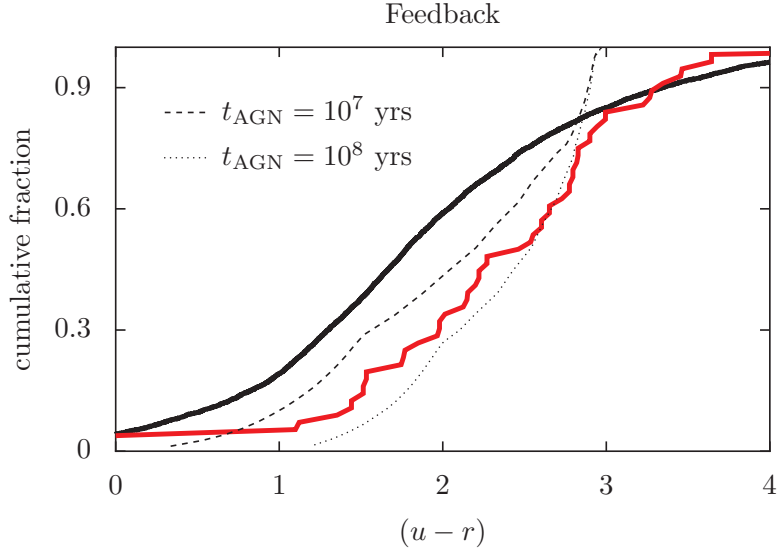}
\caption{Same as Figure~\ref{fig:outsideFit} but with AGN feedback. Colours of galaxies overrun by expanding FR-II shocks (red line) are well explained by the model.}
\label{fig:insideFit}
\end{figure}

In Figure~\ref{fig:insideFit} we show the impact of including this mode of feedback in our best-fit models. Models with AGN ages of $10-100$~Myrs do a good job of reproducing the data. These timescales are consistent with a median FR-II age of 44~Myrs in our sample, and provide strong evidence that it is the depletion of the gas reservoir by an expanding FR-II radio source that drives the colour evolution of affected galaxies.

\subsection{Recurrence timescales}
\label{sec:recurrence}

In preceding sections we have shown that powerful FR-II radio sources can (and do) affect the evolution of galaxies outside their hosts. It is important to understand how long-lasting such effects are, and whether they are important to the interplay between gas heating and cooling.

To address this question, we need to know how often powerful radio sources are triggered. Shabala \etal\/ \shortcite{SAAR08} studied a complete radio/optical sample of local galaxies, and concluded that around 1 percent of the most massive ($M_\star > 3 \times 10^{11}$~\Msun) galaxies host AGN with 1.4~GHz luminosity in excess of $10^{25}$~\WHz, which approximately corresponds to the separation luminosity between FR-Is and FR-IIs \cite{OwenLedlow94}. Given maximum FR-II lifetimes of $10-100$~Myrs, this suggests that a typical massive galaxy will launch between one and a few FR-II outbursts in a Hubble time, with the typical time between outbursts of order a few Gyrs.

The passing bow shock will both sweep up and heat the gas. The time delay before this gas is once again available for star formation is therefore equal to the sum of the cooling time $t_{\rm cool}$ and the timescale associated with reincorporation of this gas in the galaxy $t_{\rm return}$. The cooling time is a function of gas temperature and density, $t_{\rm cool} \sim 70 \left( \frac{n_e}{0.1{\rm cm^{-3}}} \right)^{-1} \left( \frac{T_{\rm gas}}{10^7 {\rm K}} \right)$~Myrs. The reincorporation timescale is typically longer than this, being comparable to the dynamical timescale for the dark matter halo hosting the cluster, $t_{\rm return} \sim \frac{0.2}{H_0 (1+z)^{3/2}} = 2.7$~Gyrs at the present epoch. These numbers suggest that FR-II outbursts in clusters can indeed prevent galaxies from forming stars despite the relatively infrequent, short-term nature of these outbursts. 

\subsection{Quiescent clusters}
\label{sec:quiescent}

The timescales of star formation suppression can also be examined observationally. For this purpose we employed a sample of 16 clusters with no detectable radio emission at 1.4~GHz, as described by Dunn \& Fabian \shortcite{DunnFabian08} and Dunn \etal\/ \shortcite{DunnEA05}. We split these clusters into regions in an attempt to mimic the process for FR-II host clusters. This is done by weighing the $R_{\rm cocoon}$ and $D_{\rm max}$ values (see Section~\ref{sec:sample}) for each cluster hosting an FR-II by the number of galaxies contained within that radius, yielding mass-weighted averages for these quantities.

\begin{figure}
\centering
  \includegraphics[height=0.45\textwidth,angle=0]{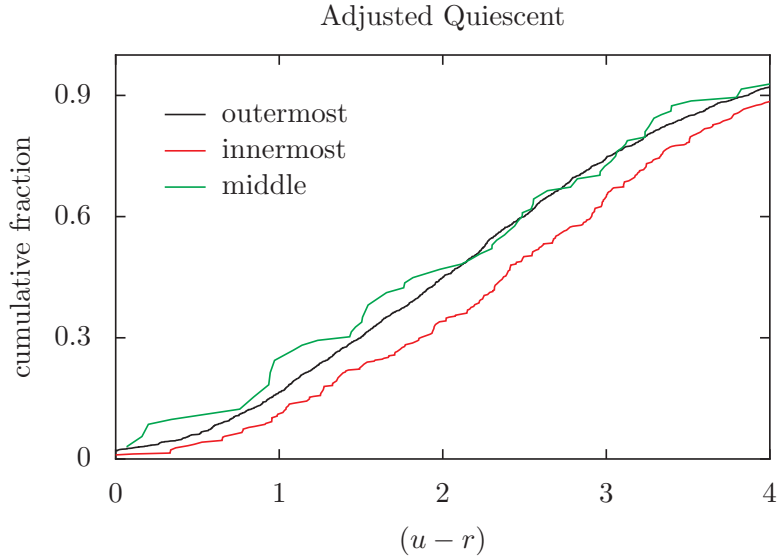}
\caption{$(u-r)$ colours for clusters with no detected 1.4~GHz radio AGN or bubbles. The three regions are chosen to match the FR-II sample, and mass corrections made. There is still a clear offset to redder colours for innermost galaxies.}
\label{fig:coloursQuiescent}
\end{figure}

Figure~\ref{fig:coloursQuiescent} shows the resultant colour distribution for the three regions. There is a clear difference between galaxies located close to cluster centre and those on the periphery, even after the usual mass corrections are performed.

\begin{figure}
\centering
  \subfigure[inside]{\includegraphics[height=0.35\textwidth,angle=0]{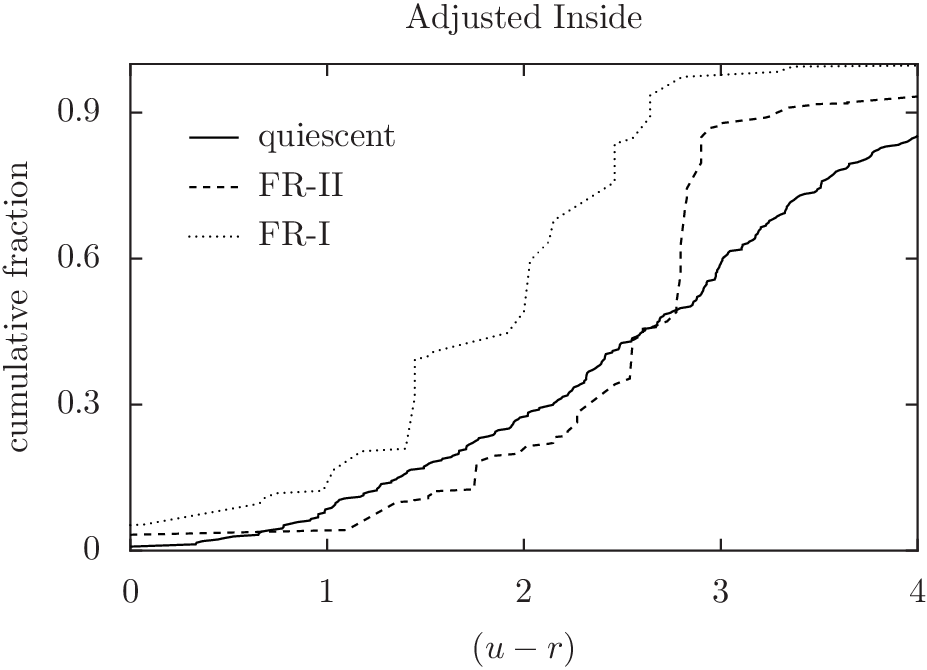}}
  \subfigure[mixed]{\includegraphics[height=0.35\textwidth,angle=0]{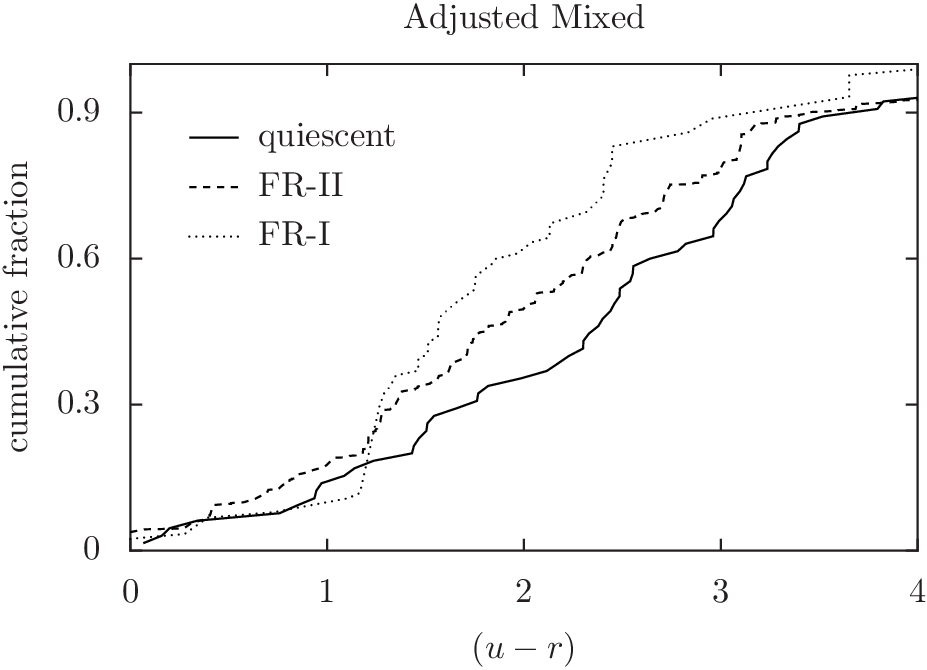}}
  \subfigure[outside]{\includegraphics[height=0.35\textwidth,angle=0]{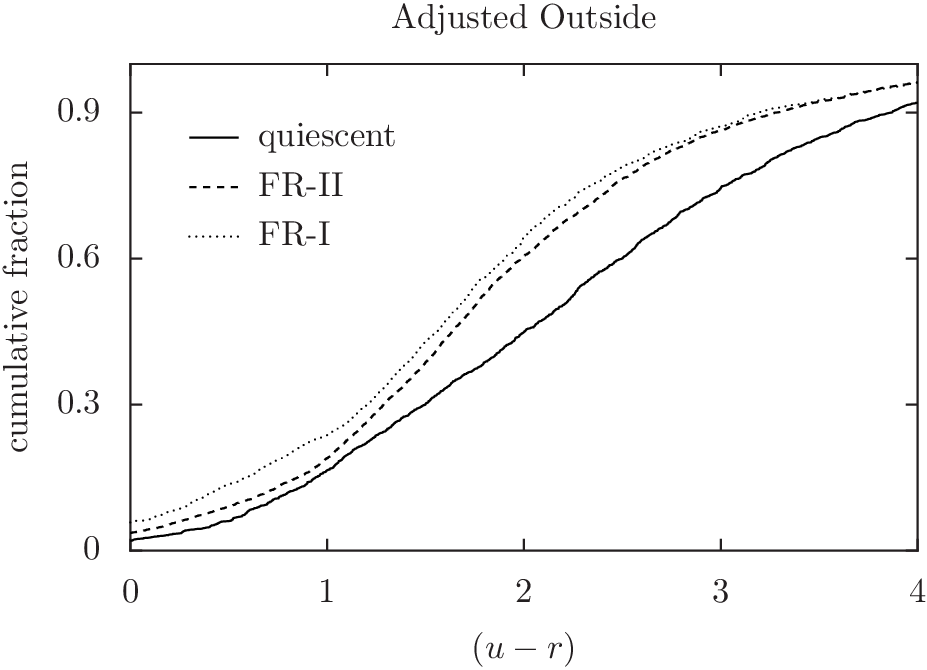}}
\caption{$(u-r)$ colours for quiescent clusters and radio AGN hosts. The AGN data have been corrected to match the mass distribution of the quiescent sample. Quiescent cluster galaxies are redder than those in both FR-I and FR-II clusters in the ``outside'' regions. In the ``inside'' region, FR-II galaxies are redder, while FR-I galaxies are bluer than the quiescent ones. These findings are consistent with a picture in which quiescent clusters are still affected by past AGN feedback, despite the absence of detectable radio emission.}
\label{fig:coloursQuiescentVsFRII}
\end{figure}

It is interesting to compare the radio source and quiescent samples. In Figure~\ref{fig:coloursQuiescentVsFRII} we plot colour distributions for the same regions in FR-I, FR-II and quiescent clusters, adjusting the mass distributions in clusters with radio sources to mimic those for the appropriate quiescent samples.

Galaxies in clusters which contain FR-I radio sources are consistently bluer than those in the quiescent cluster sample. By contrast, FR-II galaxies are marginally redder that the quiescent sample (and both much redder than the FR-I sample) in the ``inside'' region; bluer in the ``mixed'' region; and bluer still in the ``outside'' region, where the FR-I and FR-II distributions are statistically indistinguishable.

At face value, these findings are somewhat surprising. If galaxies in the quiescent cluster sample are unaffected by AGN feedback, we would expect these to have bluer colours that the FR-II sample at all radii, and perhaps even bluer colours than the FR-I galaxies if these suffer any (perhaps low-level) feedback. The results make perfect sense, however, if quiescent clusters represent a population which has undergone past AGN outbursts and has not yet regained all of the ejected/heated gas.

In this picture, an FR-II AGN is triggered, heating and ejecting gas from galaxies within the cluster. This gas cools and returns on timescale $t_{\rm cool}+t_{\rm return}$. Since FR-II lifetimes are at least an order of magnitude less than the gas return timescale (Section~\ref{sec:recurrence}), it is possible that the AGN activity terminates long before the cool gas is returned to the cluster. The cluster can then appear quiescent, but still be very much affected by the recent AGN outburst, consistent with the idea that radio sources can affect their host cluster on timescales much longer than the duration of the active phase (e.g. Fabian \etal\/ 2003, Shabala \& Alexander 2009a). Alternatively, the observable AGN signature could disappear while still undergoing feedback.

By contrast, clusters hosting radio sources must have had sufficient fuel to feed the central AGN for the past $t_{\rm AGN}$. Since $t_{\rm AGN} \ll t_{\rm return}$, the radio source triggering- and, by extension, the return of the cool gas- must have happened recently. This explains the bluer colours of galaxies in clusters which host a radio source in the ``outside'' and ``mixed'' regions. The ``inside'' region of FR-II lobes is by definition affected by feedback (since it is the volume which has definitely been overrun by the radio lobes), and is therefore devoid of gas. It is likely that these galaxies simply have not had the time to form new stars before the gas was blasted out again by the latest AGN outburst, explaining why their colours are similar to those for the quiescent sample. The marginally redder colours of galaxies in FR-II clusters are likely due to a slightly longer time since the last burst of star formation than in the quiescent sample.

\subsection{Persistence of radio emission}
\label{sec:fading}

The quiescent cluster sample was selected on the basis of a lack of 1.4~GHz radio emission. It is not immediately clear what physical condition this corresponds to. Radio sources undergo significant dynamical evolution, increasing in size and typically decreasing in luminosity after a few Myrs \cite{Alexander00}. Radio telescopes are surface brightness limited, a quantity that depends on both source size and luminosity, $\Sigma = \frac{L}{16 \pi \left( \frac{R_{\rm cocoon}}{\theta_{\rm beam}} \right)^2}$ where $\theta_{\rm beam}$ is the beam FWHM. At some point in its evolution a radio source can become dim and diffuse enough to no longer be detected in surveys. This can happen both as the AGN jet causes expansion of the radio cocoon, and after the jet terminates. Many clusters host radio bubbles (e.g. Fabian \etal\/ 2003, Forman \etal\/ 2005). These are remnants of past AGN activity, formed as the radio lobes are pinched by Kelvin-Helmholtz and Rayleigh-Taylor instabilties. The radio plasma in the bubbles is underdense, and these therefore rise buoyantly through the cluster at a speed comparable with the sound speed, sweeping up cluster gas as they do so \cite{ChurazovEA01}. If undisturbed, these bubbles can take a few Gyrs to reach cluster periphery. It is therefore important to ask is over what fraction of their lifetime the active radio AGN and bubbles are detectable in radio emission.

The electron population emitting at 1.4~GHz is subject to energy losses via synchrotron emission, adiabatic expansion, and Inverse-Compton upscattering of CMB photons. Radio bubbles are typically $>100$~Myrs old, a time when Inverse-Compton losses begin to dominate the radio luminosity. Kaiser \& Best \shortcite{KB07} give $L \propto R_{\rm cocoon}^{(-4-\beta)/3}$ in this regime, where the density profile of the ICM goes as $\rho(r) \propto r^{-\beta}$. The largest cocoon/bubble that can be detected in NVSS ($\theta_{\rm beam}=45$~arcsec, $\Sigma_{\rm min}=2.5$~mJy) is then $R_{\rm max} / {\rm kpc} = 2 \times 10^3 \left( \frac{L_{\rm 1.4GHz}}{\rm W\,Hz^{-1}} \right)^{1/2}$. Using the luminosity-size relation,

\begin{equation}
  \frac{R_{\rm max}}{R_0} = \left[ 2 \times 10^3 \left( \frac{L_{\rm 1.4GHz, 0}}{10^{26}} \right)^{1/2}  \left( \frac{R_0}{\rm kpc} \right)^{-1} \right]^{6/(10+\beta)}
\label{eqn:Rbubble}
\end{equation}
where $R_0$ and $L_{\rm 1.4GHz, 0}$ are the radius and luminosity at reference time $t_0$. Source age is related to size via an appropriate dynamical model; this age can be written as $t=t_0 \left( R(t)/R_0 \right)^x$. In the case of an active radio source, Kaiser \& Alexander \shortcite{KA97} give $x=3/(5-\beta)$, while for an inactive bubble $x=2/(6-\beta)$ \cite{KaiserCotter02}. This yields the maximum observable lifetime of

\begin{equation}
  \frac{t_{\rm max}}{t_0} = \left[ 2 \times 10^3 \left( \frac{L_{\rm 1.4GHz, 0}}{10^{26}} \right)^{1/2}  \left( \frac{R_0}{\rm kpc} \right)^{-1} \right]^{\frac{6}{x(10+\beta)}}
\label{eqn:tMax}
\end{equation}

Shabala \etal\/ \shortcite{SAAR08} derive ages and jet power for a complete sample of local radio galaxies. Here, we identify the largest, oldest sources with the jet termination phase, and therefore set $t_0=100$~Myrs, $R_0=200$~kpc, $L_{\rm 1.4GHz,0}=2 \times 10^{25}$~\WHz. This yields $t_{\rm max} = 2.5 t_0$ for the active, and $t_{\rm max}=5.8 t_0$ for the bubble phases. In other words, the radio source would become undetectable after a few hundred Myrs. This is significantly shorter than the gas return timescale, and therefore it is entirely possible that the ``quiescent'' clusters in our sample contain radio cocoons and/or bubbles undetectable in current large-scale surveys, explaining the red colours of their galaxies. The bubbles can be disrupted by a number of instabilities. However, even if they are not long-lived enough to fade and avoid detection, the long gas return timescales imply that galaxies will appear red in such clusters for an appreciable fraction of the Hubble time.

\subsection{Modes of feedback}
\label{sec:feedbackModes}

In considering FR-II and FR-I radio sources separately, we have shown that two distinct modes of AGN feedback exist. Powerful events, which we identify with the FR-II phase, are relatively rare, but can affect their environment on cosmological timescales through gas heating and expulsion. On the other hand, the less poweful FR-I events are much more frequent (by as much as two orders of magnitude; Shabala \etal\/ 2008), but cannot affect the evolution of galaxies outside their own host. This is consistent with a picture in which FR-I sources start out as FR-IIs and are simply disrupted within the dense galaxy core. AGN counts in the local volume are dominated by these low-luminosity events, and are consistent with the AGN being triggered by cooling of gas out of the hot halo \cite{BestEA05b,SAAR08}. Recently Shabala \& Alexander \shortcite{SA09b} have shown that this low-level mode of feedback is sufficient to stop star formation in massive galaxies at late times and match the local stellar mass function.

Observations of the AGN fraction are insensitive to the duration of the active and quiescent phases of radio activity, instead only providing information on the ratio $t_{\rm active}/(t_{\rm active}+t_{\rm quiescent})$. As discussed above, both these timescales will be significantly longer for FR-II events than for FR-Is. Galaxy colours, on the other hand, are sensitive to the {\it absolute} timescales, rather than just this ratio. It appears likely that the few very red galaxies in Figure~\ref{fig:insideFit} that cannot be explained by our simple model may reside in clusters which have hosted more than one powerful FR-II outburst. In this scenario, if re-triggering of the AGN in the FR-II host is facilitated by a sudden influx of cold gas (as would happen, for example, in a gas-rich minor merger), it can proceed without catastrophic cooling in other galaxies within the cluster. Repeated gas ejection outbursts are therefore possible without associated star formation. On the other hand, FR-I mode of feedback is quite different in that only the host galaxy is affected by the radio source. This means that gas must be supplied to the galaxy for another AGN outburst to take place, and the outburst will therefore usually be accompanied by circumnuclear star formation. These galaxies will typically appear bluer. Kaviraj \etal\/ \shortcite{KavirajEA11} find that gas ejection is required to explain the colour evolution of early-type galaxies from the blue cloud to the red sequence. This could be provided by either of the two modes of feedback discussed here. However, it is difficult to envisage a scenario in which small-scale AGN outbursts can explain the colours of the reddest observed galaxies. Galaxies recently overrun by FR-II radio sources show no correlation between the radio source age and $(u-r)$ colour, suggesting that the colours are only sensitive to AGN feedback on timescales exceeding typical radio source lifetimes of $< \sim 100$~Myrs.

\section{Summary}
\label{sec:summary}

We present an analysis of galaxy colours around FR-I and FR-II type radio sources. We find that the less powerful FR-I radio sources cannot affect the colours of galaxies near AGN hosts; while galaxies overrun by expanding FR-II radio sources exhibit redder colours that are consistent with truncation of star formation following the passage of a radio source-driven bow-shock. We compare the FR-I and FR-II samples with galaxies located in clusters that show no evidence of current AGN activity, and find that these quiescent clusters are redder than expected for a secular evolution scenario. Galaxy colours can be affected on timescales significantly exceeding the detectable AGN lifetime, and rare powerful AGN events thus play an important role in the colour evolution of local galaxies.

\section*{Acknowledgements}

We thank the referee, Elaine Sadler, for insightful, thorough and constructive comments that have undoubtedly improved the manuscript. SSS thanks New College, Oxford for a research fellowship and the BIPAC institute at Oxford for support. SK acknowledges a Research Fellowship from the Royal Commission for the Exhibition of 1851, an Imperial College Junior Research Fellowship, a Senior Research Fellowship form Worcester College, Oxford and support from the BIPAC institute. SSS thanks Anna Scaife for assistance with the radio data, and Dominic Ford for creating PyXPlot.

%SDSS thanks.
Funding for the SDSS and SDSS-II has been provided by the Alfred P. Sloan Foundation, the Participating Institutions, the National Science Foundation, the US Department of Energy, the National Aeronautics and Space Administration, the Japanese Monbukagakusho, the Max Planck Society and the Higher Education Funding Council for England. The SDSS web site is http://www.sdss.org.

The SDSS is managed by the Astrophysical Research Consortium for the Participating Institutions. The Participating Institutions are the American Museum of Natural History, Astrophysical Institute Potsdam, University of Basel, University of Cambridge, Case Western Reserve University, University of Chicago, Drexel University, Fermilab, the Institute for Advanced Study, the Japan Participation Group, Johns Hopkins University, the Joint Institute for Nuclear Astrophysics, the Kavli Institute for Particle Astrophysics and Cosmology, the Korean Scientist Group, the Chinese Academy of Sciences (LAMOST), Los Alamos National Laboratory, the Max-Planck-Institute for Astronomy (MPIA), the Max-Planck-Institute for Astrophysics (MPA), New Mexico State University, Ohio State University, University of Pittsburgh, University of Portsmouth, Princeton University, the United States Naval Observatory and the University of Washington.

%FIRST thanks. NVSS thanks.
The NVSS and FIRST surveys were carried out using the National Radio Astronomy Observatory VLA. The National Radio Astronomy Observatory is a facility of the National Science Foundation operated under cooperative agreement by Associated Universities, Inc.

\end{document}